# Numerical model of the inhomogeneous scattering by the human lens

**ALEXANDER CUADRADO[1], LUIS MIGUEL SANCHEZ-BREA[2], FRANCISCO JOSE TORCAL-MILLA[3], JUAN ANTONIO QUIROGA[2,4] AND JOSE ANTONIO GOMEZ-PEDRERO[1,*]**

[1]*Applied Optics Complutense Group, Optics Department, Universidad Complutense de Madrid, Facultad de Óptica y Optometría, C/ Arcos de Jalón, 118, 28037, Madrid, SPAIN.*
[2]*Applied Optics Complutense Group, Optics Department, Universidad Complutense de Madrid, Facultad de Ciencias Físicas, Ciudad Universitaria, Plaza de las Ciencias, 1, 28040, Madrid, SPAIN.*
[3]*Applied Physics Department, Universidad de Zaragoza, C/ Pedro Cerbuna, 12, 50009, Zaragoza, SPAIN.*
*4 Indizen Optical Technologies, .*
*[*jagomezp@ucm.es](*jagomezp@ucm.es)*

**Abstract:** We present a numerical model for characterizing the scattering properties of the human lens. We have analyzed the scattering properties of two main scattering particles actually described in the literature through FEM simulations. From the data obtained we have modified a Monte Carlo's bulk scattering algorithm for computing ray scattering in non-sequential ray tracing. Finally we have implemented the bulk scattering algorithm in a layered human lens model in order to calculate the scattering properties of the whole lens. We have tested our algorithm by simulating the classic experiment carried out by Van der Berg *et al* for measuring "in vitro" the angular distribution of scattered light by the human lens. The results shows the ability of our model for simulating accurately the scattering properties of the human lens.

## References and links

## 1. Introduction

The human lens is a remarkable biological structure. Its mission is to focus the light refracted at the cornea into the retina. The human lens (sometimes we will simply refer to it as lens for the sake of clarity) is a layered structure formed by cells which contain an intracellular medium composed by a solution of proteins (mainly of the crystallin protein family) in water

[1]. The concentration of the proteins is higher in the lens core than in the cortex which explains the refractive index growing between the periphery and the core as the refractive index of the lens depends on the concentration of crystallin proteins in water. Therefore, the human lens acts as a gradient index (GRIN) lens. This fact has been demonstrated both in animals [2] and humans [3]. The fiber cells provide the mechanical properties of the lens which is rigid enough to maintain its biconvex lens-like shape but also flexible to allow accommodation. On the other hand, the intracellular media provides the lens transparency and refractive index inhomogeneity of the human eye lens. It is important to point out that, contrary to what happens with other tissues, the normal human lens does not present structures that might produce high amounts of scattered light, such as blood vessels, cell nuclei, organelles, or other potential scattering structures. Moreover, in normal conditions, there is an important index matching between the cell walls and the intracellular medium, particularly at the lens core, which reduces light scattering and/or diffraction at these boundaries [1].

Despite the mechanisms described before, some scattering is always present in the human lens. In normal eyes, scattering only manifests itself when the eye is illuminated with high amounts of energy. Even for normal eyes, ocular scattering gives rise to the so called "straylight" defined by Van der Berg [4] which is related to the peripheral part of the ocular PSF and to the physiological phenomenon of glare [4]. The amount of light scattered by the eye lens is greater for aging individuals, particularly when cataracts are present. In this case, the individual will experience some of the different symptoms associated with this condition such as loss of visual acuity and contrast, more frequent appearing of glare, etc. being all of them related with an increasing light scattering within the lens.

Light scattering at the human lens has been thoroughly studied both theoretical- and experimentally by many authors. Bettelheim *et al* [5] measured "in vitro" the angular distribution of scattered light by sections (of around 10 µm thick and perpendicular to the optical axis) of the crystalline of different donors. They fitted their results to a theoretical model based on random fluctuations of density and orientation of scattering particles, and they concluded that the eye lens scattering was due to a mixture of two scattering particles: a spherical protein cluster with a radius between 130 and 450 nm and cylindrical cytoskeletal bodies with a length comprised between 600 and 2600 nm [5]. In any case, a strong dependence between the model parameters and the spatial location of the section within the crystalline was observed. Subsequent works expanded this model by studying the dependence with age [6] and eyes presenting different kinds and degrees of cataracts [7]. Notice that all of these experiments have been performed "ex vivo" that is by extracting the lens after the decease of the donor, and we have to keep in mind that scattering properties of extracted lenses ("ex vivo") may not be the same as that of lenses in a living eye ("in vivo").

Van der Berg and Ijspeert measured "in vitro" the light scattering of whole lenses [8] from several donors of different ages, some of them clear and some of them presenting cataracts. In most cases, the fitting of the resulting scattered intensity against the polar angle θ shown an inverse squared dependence. In a later work [9], Van der Berg and Spekreijse were able to measure "in vitro" the light scattered by different parts of the human lens using slit illumination, avoiding, in this way, the need for slicing the lens. The result of this analysis was a phenomenological model which explains the lens scattering through the contribution of three sources. The first source is the Rayleigh scattering produced by small particles with a radius around 10 nm, closely related to that of the α-crystalline protein aggregates [9]. The second source is the non-Rayleigh scattering produced by relatively big (700 nm radius) "effective" particles. Finally, the third source described is the surface scattering at the zones of discontinuity (Wasserspalten) corresponding to the different layers of the human lens [4,9]. It is noteworthy that the phenomenological model presented by Van der Berg and Spekreijse shows a spatial dependence due to the inhomogeneity of the lens.

The exact nature of the large scattering particles within the human lens as multilamellar bodies was reported by Gilliland *et al* [10] and thoroughly analyzed by Costello *et al* [11,12].

These multilamellar bodies, or vesicles as we will call them in the following, are structures formed by a lipid shell that surrounds a body with the same composition of water and proteins than that of the human lens intercellular medium. The mean size of these particles are around 2.13 microns in diameter. As it is shown by Costello *et al* [11] and Mendez.Aguilar *et al* [13], the index matching between the vesicle shell, the inner body and the external medium makes these vesicles highly efficient scatterers with a relatively high cross-section. In all of these works, the concentration of the vesicles within the human lens is deemed as constant but not the size, which is distributed along the lens following a random Gaussian distribution with a mean diameter of 2.13 µm and a standard deviation of 0.64 µm [11] .

Regarding the nature of the small size scatterers, we will not take into account in our model the soluble alpha-crystallin aggregates reported by Van den Berg and Spekreise [9] because of their small size and the fact that they are so closely packed (particularly in the core of the lens), that they cannot be considered as independent scatterers and they contribute more to raise the refractive index of the lens core [14]. Instead, we will focus our attention in the insoluble high molecular weight (HMW) aggregates that are formed by gamma crystallin denaturalization [15][16]. These aggregates were first reported in the 1970's and they have been thoroughly studied ever since [17][18]. In this context, it was Benedek [19] who first predicted theoretically that light scattering in cataractous eyes could be due to protein aggregates with a molecular weight of around $50 \times 10^6$ g/mol (with corresponding estimated diameter of 50 nm). This prediction was confirmed experimentally by several authors [20][21] using different techniques. Nowadays there is a good knowledge of the biochemical processes that lead to the formation of HMW aggregates [22][23] basically through the inhibition of the chaperone activity of the alpha-crystallin and denaturalization of gamma-crystallin which becomes susceptible to aggregate forming structures such as amyloid fibers [24] or protein condensates [25]. These structures have been measured with different techniques such as electron microscopy or magnetic resonance [24][25]. From these measurements, diameters ranging from 20 nm to 30 nm [24][25] have been reported. An additional cause of light scattering in the eye are the zones of discontinuity but their effect is greater for the back-scattered light, so it will not be considered in our model.

In this work, we have developed a model for the numerical computation of the scattering of the eye lens. According to the literature, we have considered that light scattering in the human lens is produced by a mixture of small (clusters) and large (vesicles) particles and we have performed a comprehensive study of the scattering properties of each kind of particle. We have also taken into account the inhomogeneous nature of the human lens which is manifested in the spatial variation of the protein concentration within the lens. As a consequence of this variation of the protein concentration, both the refractive index and some scattering parameters also vary spatially. Although our model has been developed using a particular ray tracing software (Zemax OpticStudio$^{TM}$) it could also have been implemented in other ray tracing programs or even as a stand-alone application using a general programming language such as C++, C#, Java or Python. In order to test our model, we have made a simulation of the human lens scattering experiment described in reference [8] and we have performed a non-linear fit with the experimental results described in this paper in order to find the optimum values of the parameters of our model for both a normal and a cataractous eye.

The paper is organized as follows: in Section 2 we describe the basis of our model, including a thorough FEM electromagnetic study of the scattering properties of clusters and vesicles, and the implementation of a mixture model of those scattering particles (clusters and vesicles) for computing the bulk scattering of a ray using the Monte-Carlo technique. In Section 3, we give the results of the numerical simulation of the experiment reported in reference [8], in order to show the ability of our model to describe properly the light scattering by the human lens. Finally, conclusions are drawn to end the paper.

## 2. Inhomogeneous scattering model of the human eye

*2.1 Theoretical foundations*

A healthy human lens acts like a high transmittance GRIN lens presenting a spatial variation of the refractive index. This spatial variation is due to the different concentration of protein within the human lens [1]. We can stablish a relationship between this variable protein concentration and the refractive index of the eye, $n$, using a simple mixing model

$$n(c) = (1-c)n_w + cn_p, \qquad (1)$$

where $c$ is the protein concentration, $n_w = 1.33$ the refractive index value of water, and $n_p = 1.62$ the refractive index of the protein according with data obtained in different experiments [2,3].

Typically, a mix of heterogeneous particles is found across the intercellular mediums. If the size of these particles remains small and their concentration low, the resulting scattering is negligible [26,27]. However, some factors, such as age, UV radiation, and local thermal variations, trigger the increase in particle size and number, generating a notably enhancement of scattering effects [5].

The scattering of a particle is strongly related to its geometry, the light wavelength and the relationship between the refractive index of particle and its surrounding medium. In this study we neglect the absorption of the scatterers within the lens. Within this approach, two scattering magnitudes become relevant when analyzing the light-particle interaction. In our model, these magnitudes depend on the local concentration of protein within the lens. The first one is the scattering cross-section, $\sigma_s(c)$, which describes the probability function of incident electromagnetic radiation being scattered by a particle, calculated as [26,28-32]:

$$\sigma_s(c) = \frac{1}{I_o} \oiint \mathbf{S}_s(c) \mathbf{n} \, dA, \qquad (2)$$

where $\mathbf{S}_s(c)$ is the Poynting vector of the scattered radiation, $\mathbf{n}$ the normal vector, and $I_o$ is the incoming irradiance. The integration domain corresponds to the area of an imaginary sphere which surrounds completely the scattering particle. The second parameter of interest, the coefficient of asymmetry, describes the anisotropy of the scattered radiation, and it is computed as [32]:

$$g(c) = \frac{1}{\sigma_s I_0} \oiint \mathbf{S}_s(c) \mathbf{n} \cos\theta \, dA, \qquad (3)$$

where the integration is again carried out over the surface of an imaginary sphere that surrounds the particle, being $I_0$ the incoming irradiance, $\sigma_s$ the scattering cross section, $\mathbf{S}_s(c)$ the Poynting vector of the scattered radiation at a given point of the surface of the sphere, $\mathbf{n}$ the normal vector at the same point, and θ the polar angle defined as the angle formed by the direction vector of the incoming radiation and the normal vector $\mathbf{n}$. In our study we have used the far-field pattern for computing both the cross section and the asymmetry coefficient according to [28,32].

We assume that the concentration of the scattering particles is low enough to disregard interference effects between scatterers. Moreover, as a first approximation, we also consider a medium made of identical particles. Then, the scattering coefficient of our medium is given as:

$$\mu_s(c) = \sigma_s(c) N, \qquad (4)$$

being $N$ the volumetric concentration (particles per unit of volume) of the scattering particles. We can define a scattering mean-free-path, $l_s(c) = \mu_s(c)^{-1}$, as the distance travelled by light between two scattering events. In principle the scattering phenomena is described by the scattering coefficient and mean free path. However, a non-absorbing turbid media with particles presenting high asymmetry, such as the human lens, is better described [31] by defining the reduced scattering coefficient: $\mu'_s(c) = \mu_s(c)(1 - g(c))$ and the effective mean-free-path as: $l'_s(c) = \mu'_s(c)^{-1}$.

Turbid media, particularly, biological tissues, show scattering probability distributions strongly related to $g$. The Henyey-Greenstein scattering phase function includes the anisotropy parameter to calculate the phase function, $p(\theta)$ as [27,31]:

$$p(\theta) = \frac{1-g^2}{4\pi(1+g^2-2g\cos\theta)}, \qquad (5)$$

where $\theta$ is the polar angle. Notice that the phase function depends indirectly on the concentration through the asymmetry coefficient $g$. Next, we will show the results obtained after computing the scattering parameters of different scatterers present in the human lens using finite element methods (FEM).

*2.2 Scatterers in the human lens*

All particle simulations considered in this paper have been evaluated using a FEM software, COMSOL™ Multiphysics 5.2. In every simulation, we use an electromagnetic radiation with wavelength of 550 nm, according to the maximum sensitivity of human eye. Also, we have taken in all cases the amplitude of the incident electric field amplitude as $E_0 = 1$ V/m.

Protein Clusters

We can consider the crystalline protein as a spherical particle with a radius about 2 nm. When these proteins agglomerate, they form protein clusters that can be considered as a greater particle which radius up to 10 nm for normal eyes [24]. According with the optimal close sphere packing fraction whose value is around 75%, these new dispersers are not composed by pure proteins. Thus, the refractive index of these clusters decreases until a value of $n_c \approx 1.55$ from the value of $n_p = 1.62$ which presents the bulk protein. For normal eyes, the scattering effect of these clusters is small enough to not impair the transparency of human lens [34]. However, in certain conditions (crystalline aging, temperature increment, etc.), the cluster radius increases up to 20 nm [24] or greater (around 50 nm) [25]. In this case the radiation scattered by these clusters rises significantly according to Mie theory.

Fig. 1 shows the near- and far-field patterns of the scattered radiation, the scattering cross section, and the asymmetry coefficient, as a function of the radius of the cluster and the protein concentration of the surrounding medium. Protein cluster shows a typical dipolar near field pattern for a particle with 25 nm radius. However, we get a micro-jet response for cluster radius of 150 nm. The scattering cross section increases when the radius becoming larger as shown in Fig 1d), and it also depends on the protein concentration of the surrounding medium. When the protein concentration is low, the refractive index jump between the surrounding medium and the protein cluster increases, raising the value of the scattering cross section. The anisotropy of protein clusters can be studied considering the far-field pattern of the scattered radiation. Fig.1 a), b) and c) shows this far-field pattern (see insets). For low cluster radius, we see a dipolar behavior, being the anisotropy of cluster particles close to $g = 0$. However, when the radius increases, the anisotropy grows up towards $g = 1$. On the other hand, anisotropy does not dependent strongly with the protein concentration value (see Fig. 1e). By interpolating the data depicted in Fig 1 we can obtain the value of the scattering cross section and asymmetry coefficient for any value of the protein concentration and cluster radius.

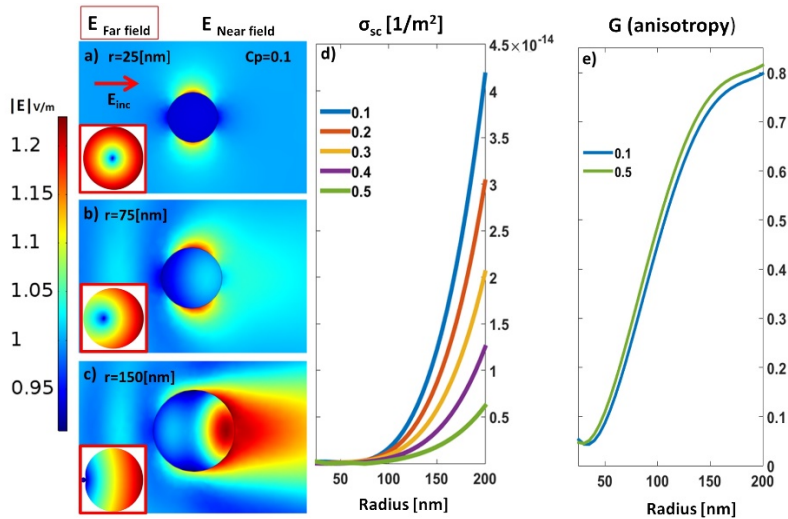

Fig. 1. Near field distribution of the scattered electric vector modulus for a spherical cluster (refractive index $n_c = 1.55$), embedded in a medium with a protein concentration c=0.1 (corresponding to a refractive index $n = 1.359$), for cluster radius of a) 25, b) 75 and c) 150 nm, respectively. The arrow indicates the direction of the incident field. In the insets we have represented the far field distribution for the same cluster radius. d) Scattering cross section of a cluster as a function of the cluster radius, for different values of the protein concentration of the surrounding medium. e) Coefficient of asymmetry, g, as a function of the cluster radius for two extreme values ($c = 0.1, 0.5$) of protein concentration.

Vesicles or multillamelar bodies.

The fiber cells that structure the human lens show different shapes depending on when they were generated. Thus, the shape of a fiber cell grown up during the human fetal period shows a greater uniformity than a structure generated at larger period of life [10]. Thorough the lens, breaks at the cells' walls, [11], generate a type of bulges known as multilamellar bodies or vesicles [11, 13]. To model these vesicles, we consider a particle, which average diameter around 2.3 μm, having a lipid shell with a thickness of 50 nm that shows a refractive index around 1.55 [11,13]. Following Costello *et al* [11], we have considered that the refractive index of the inner medium of the vesicle is similar to the surrounding one, but presenting a 6% increment in the refractive index. Fig 2 shows the modulus of the electric vector corresponding to the near- and far field patterns, the scattering cross section and the asymmetry coefficient for different protein concentration of the surrounding medium and particle size.

As shown in Fig. 2, the vesicles display an electromagnetic near-field pattern that can be compared with that of a micro-lens. Due to their greater size, compared with clusters, the scattering cross section of vesicles is larger than that of protein clusters, making them more efficient scatterers. Indeed, there is a strong dependence of the scattering parameters with the diameter as displayed in Fig. 2. Notice that for a vesicle with radius 1.4 μm, we get a scattering cross section of around $1.75 \cdot 10^{-11}$ m$^{-2}$. Divided by the area of the vesicle we get a scattering efficiency of 2.84, which is of the same magnitude order of that obtained by Costello et al [41]. However as the ratio between the inner refractive index and that of the external medium is fixed, there is almost no dependence with protein concentration as shown in Fig, 2c.

The electromagnetic far-field pattern of the vesicles shows a strong forward scattering distribution -see red box insets of Fig. 2a and 2b)-. This effect corresponds with an anisotropy value around $g = 0.9$, a value that it is only obtained for large protein clusters ($r_c = 200$ nm). The anisotropy grows when the protein concentration becomes larger.

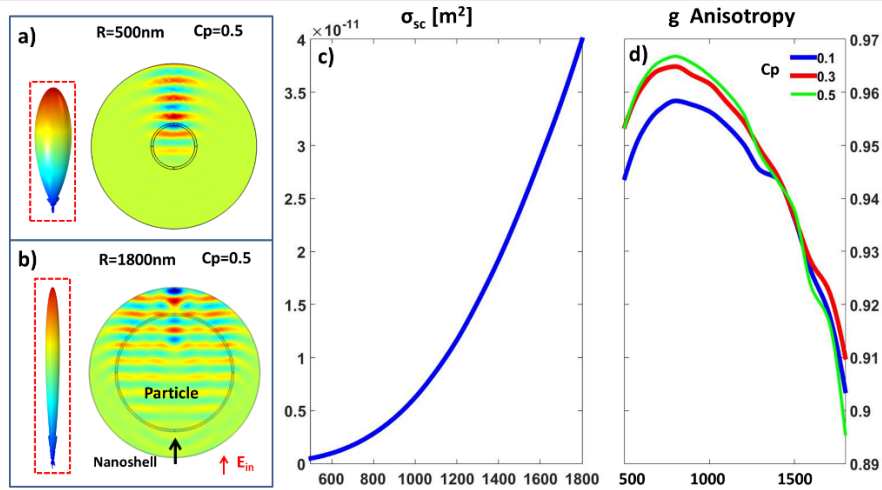

Fig 2: Near field modulus of the electric vector corresponding to a vesicle of a) 500 nm radius, b) 1800 nm radius, both with a shell 50 nm thick computed for protein concentrations of $c = 0.5$. The red arrow indicates the direction of the incident field. The far field pattern modulus is also displayed (see instets). c) Scattering cross section of vesicles as a function of the vesicle radius. d) Asymmetry coefficient as a function of vesicle radius for different protein concentration.

As with clusters, we can compute the values of the scattering cross section and asymmetry coefficient by interpolating the data resulting from the FEM simulation.

### 2.3 Monte Carlo's bulk scattering algorithm for human lens particles

Once we have computed the properties of the different scattering particles within the human lens (clusters and vesicles) we can now proceed with the implementation of a computer bulk scattering algorithm. In this work we have modified the Monte Carlo's bulk scattering algorithm provided by Zemax OpticSudio$^{TM}$ (ZOS in the following), although our method can be easily extrapolated to other ray tracing programs or programming languages. More details on the ZOS bulk scattering algorithms can be found in reference [37], while its implementation in an ocular model is reported in reference [38]. Our departing point is the algorithm that works with the Henyey-Greenstein probability distribution (5), described in reference [39].

In order to modify this algorithm for the scattering particles of the human eye we must take into account some considerations. First, we are dealing with an inhomogeneous medium whose refractive index depend on the local concentration of protein. Second, the scattering parameters of both clusters and vesicles depend on the external medium refractive index (through the protein concentration). Finally, the size (and volumetric concentration in the case of clusters) of both scattering particles varies also within the lens. In the case of clusters we will model this variation in size and concentration after Siew et al [SIEW], while for vesicles we will consider a random variation of size with a constant concentration following Costello et al [11].

As we will see with more detail in the next section, we have modeled the lens as a layered medium composed by several layers of different refractive index but being each layer a homogeneous one. In this way, we may kept the same homogeneous Monte Carlo's bulk scattering algorithm for all the layers of the lens.

The parameters of the bulk scattering algorithm are the cluster radius $r_c$ expressed in nm, the mean distance between clusters $d$, also in nanometers, and the vesicle concentration $N$, measured in particles/mm$^3$. For normal eyes, the values of these parameters are around $r_c = 20$ nm, $d = 1100$ nm and $N = 500$ mm$^{-3}$. For cataractous eyes, $r_c$ and $N$ are usually greater and $d$ lower than these reference values. As said before, each layer of the lens may have different values of these parameters in order to model the non-homogeneous scattering properties of the human lens. The bulk scattering algorithm can be described by the following steps:

1. For each ray passing through the bulk medium of the layer, the algorithm computes a random vesicle diameter using a Gaussian distribution with mean 2.13 μm and standard deviation of 0.64 μm [11]. This is an approximation to the fact that the ray encounters vesicles with random sizes when it propagates through the lens. We also compute the volumetric cluster concentration as $N_c = \sqrt{2} \cdot d^{-3}$, being $d$ the mean distance between clusters.
2. Next, the local protein concentration is determined from the value of the local refractive index and the simple mixture model given by equation (1) so

$$c = \frac{n - n_w}{n_p - n_w}, \tag{6}$$

Where, as we said before, $n$ is the refractive index of the layer, $n_p = 1.62$ the refractive index of the protein and $n_w = 1.33$ that of the water. From the protein concentration, and the size of the scattering particles, we compute the scattering coefficients $\sigma_s$ and $g$ for both clusters and vesicles using the results of the electromagnetic model depicted in Figs. 1 and 2.
3. Afterward, we compute, for each scattering particle, the mean free path, $l'_s$, as the inverse of the reduced scattering coefficient $\mu'_s$ which, in turn, depends on the asymmetry coefficient, $g$, the scattering cross section, $\sigma_s$, and the volumetric concentration of the scatterers, $N$, see equation (4).
4. Next, we have to manage the mixture of two different scatters in the Monte Carlo's algorithm. To do so, we have followed the procedure described in reference [39]. According to this work, when we have a mixture of two kinds of non-interacting scattering particles, clusters and vesicles in our case, characterized by their free mean paths $l'_{sc}$ and $l'_{sv}$, respectively, the probability that the scattering particle is a cluster is $p_c = l'_{sc}/(l'_{sc} + l'_{sv})$. Conversely, the probability that the scattering particle is a vesicle is $p_v \equiv 1 - p_c = l'_{sv}/(l'_{sc} + l'_{sv})$. In these conditions, for each ray that passes through the medium, we can determine whether a cluster or a vesicle could scatter the ray by extracting a random number $\xi \in [0,1]$. If $\xi \leq p_c$ then the ray could be scattered by a cluster, otherwise it could be scattered by a vesicle.
5. Finally, the bulk scattering algorithm will determine whether or not the ray is actually scattered, and, if so, it computes the direction of the scattered ray using the Henyey-Greenstein probability distribution (5). To do this, the algorithm uses the values of the mean free path and the asymmetry coefficient as stated in references [27][37].
6. The procedure is repeated for each ray traced through the medium to get the final distribution of the scattered light. In our simulations we have used a number of rays comprised between $10^7$ and $10^8$.

To implement this algorithm we have programed a DLL in C language which can be used by the bulk scattering feature of ZOS.

## 2.4 Modeling human lens as a layered medium

The final step of our model is the modeling of the eye lens as a layered GRIN medium. To do so we have used the capabilities of the non-sequential ray tracing mode of ZOS. In this mode it is possible to define nested volumes, so we can simulate a layered GRIN media by defining a set of nested volumes each one with a different refractive index. The reader can found further information and examples in reference [40]. Obviously, in a GRIN layered medium defined as a set of nested volumes, the surfaces that delimit these volumes are iso-refractive index surfaces.

Therefore, we need a model for the refractive index distribution of the lens so we can find the iso-refractive index surfaces. In the literature, there are several refractive index models available and we have selected the model proposed by Navarro et al [41]. We have done so because this model takes into account the age as a parameter, which is very well suited for studying cataracts associated with age, and because its great optical and morphological accuracy. In Navarro's model the refractive index is given by the following function

$$n_{ant}(z,\omega) = n_0 + \delta_n \left(1 - \frac{1}{f_{ant}}\left(\frac{z^2 - 2\Delta_{ant}z}{a_{ant}^2} + \epsilon_{ant}\frac{\omega^2}{b_{ant}^2}\right)\right)^p \quad for\ (z_{ant},\omega_{ant}) \leq (z,\omega) \leq (z_i,\omega_i).$$

$$n_{pos}(z,\omega) = n_0 + \delta_n \left(\frac{1}{f_{pos}}\left(\frac{z^2 - t_{ant}^2 - 2\Delta_{pos}(z-t_{ant})}{a_{pos}^2} + \epsilon_{pos}\frac{\omega^2}{b_{pos}^2}\right)\right)^p \quad for\ (z_i,\omega_i) \leq (z,\omega) \leq (z_{pos},\omega_{pos})$$

$$, \quad (7)$$

with $\omega^2 = x^2 + y^2$ being the radial coordinate, $z$ the axial coordinate, $n_0$ the maximum central index value and $\delta_n$, $f_{ant}$, $\Delta_{ant}$, $a_{ant}$, $b_{ant}$, $\epsilon_{ant}$, $t_{ant}$, $f_{pos}$, $\Delta_{pos}$, $a_{pos}$, $b_{pos}$, $\epsilon_{pos}$ and $p$ are constants which depend on the age and accommodation. These constant can be computed from the data tabulated in table 1 of reference [41]. In this work we have used the constants corresponding to a young subject (24 year old) and al elderly one (80 year old) with no accommodation (we will see in next section why we have chosen this ages). In equation (7) $(z_{ant},\omega_{ant})$, $(z_i,\omega_i)$, and $(z_{pos},\omega_{pos})$ are the coordinates that describes the anterior, intermediate and posterior surface of the lens (see details on reference [41]).

By imposing the condition: $n_{ant}(\omega,z) - n = 0$, where $n$ is a value of the refractive index, we get an iso-indical surface located in the anterior portion of the lens. Similarly, the solutions of the equation $n_{pos}(\omega,z) - n = 0$ will define an iso-indical surface located at the rear of the lens. Solving these equations for different values of $n$ we get a family of iso-indical surfaces $z = f(\omega;n)$ that define the different layers of the human lens. According to Navarro's model these iso-indical surfaces are conicoids described by the general equation

$$z(\omega) = z_0 + \frac{\kappa \omega^2}{1 + \sqrt{1 - (1+c)\kappa^2\omega^2}}, \quad (8)$$

Where $z_0$ is the location of the surface vertex measured from the vertex of the lens, $\kappa = R^{-1}$ is the vertex curvature, and $c$ the conic constant., each one given by equation (8), and a center thickness, $t_0$. All of these parameters depend on whether the surface belongs to the anterior or posterior zone of the lens and on the value of the refractive index [41].

In this work, we have modeled the human lens with 11 layers corresponding to the set $n \in \{1.362, 1.366, 1.370, 1.374, 1.378, 1.382, 1.386, 1.390, 1.394, 1.398, 1.402\}$. Each of this layer is a lenticular volume defined by two conicoid surfaces, so our lens model consists of a set of nested aspheric lenses. In Fig. 3 we can see the layered model of the human lens for a 24 and an 80 year old subject. Notice the significant differences in the lens morphology and refractive index distribution for these two ages.

One of the basic assumptions of our model is that each layer contains scatterers (clusters and vesicles) with different sizes and concentrations. This is in agreement with the experiments of Siew *et al* [6] and Van der Berg and Speckrijse [9], which show that the scattering particles are not homogeneously distributed through the human lens so that their mean size and concentration vary along the axial direction *z*. This is particularly true for clusters [6] while for vesicles it is usually assumed a uniform concentration along the eye lens [11] with a randomly distributed diameter. In our layered model of the human lens, we can change independently for each layer the parameters that define the size of clusters and the concentration of both clusters and vesicles (note that, as stated before, the size of vesicle is selected randomly using a

Gaussian distribution by the Monte Carlo algorithm), although we keep this later parameter equal for all the layers as done in previous works [11].

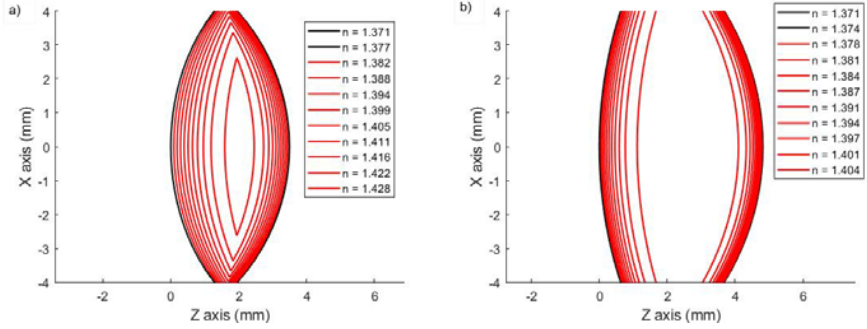

Fig 3 Layered model for the human lens based in Navarro's index model for: a) a 24 year old subject and b) a 80 year old subject. Notice the difference in curvature and thickness of both lenses and the different distribution of the iso-indical curves.

In order to compute the cluster radius and distance for each layer, we have fitted the data provided by Siew *et al* [6], obtaining the following spatial dependence for the cluster radius

$$r_c(z) = r_0(1 + b_1 \cos wz + b_2 \sin wz), \tag{9}$$

where $r_o$ is an "average" cluster radius, $z$ is the axial distance measured from the apex of human lens, and the remaining parameters are $b_1 = -0.0059$, $b_2 = 0.0865$ and $w = 2.74$ rad/mm. To implement this function in our model we select first a value for $r_0$ and then we compute the cluster radius for each layer by evaluating equation (9) at the locations given by the values of $z_0$ corresponding to the vertex of the iso-index surfaces that define the layered model of the lens.

We have proceed in a similar way for the cluster distance *d*. Indeed we have selected this parameter for characterizing the cluster concentration because its spatial dependence is given by Siew *et al* [6]. After fitting the experimental data reported by these authors, we have found that the spatial dependence of the cluster distance is given by the following function

$$d(z) = d_0(1 + c_1 z + c_2 z^2 + c_3 z^3), \tag{10}$$

Being $d_0$ the cluster distance at $z = 0$ (the front vertex of the lens), and $c_1 = -0.742$ mm$^{-1}$, $c_2 = 0.7$ mm$^{-2}$, and $c_3 = -0.225$ mm$^{-3}$. Therefore by selecting a value for $d_0$ and substituting the values of $z_0$, in equation (10) we get the cluster distance for each layer.

Using equations (9) and (10) we have effectively reduced the parameters of the scattering particles in our model to three: $r_0$, $d_0$ and $N_v$, being this latter parameter is the volumetric concentration of vesicles.

To summarize, we have modeled the eye lens as a set of nested volumes, each one of them an aspheric lens whose anterior and posterior surfaces are iso-indical surfaces of Navarro's gradient index model. For each layer, the concentration and size of clusters vary according to equations (9) and (10) but the vesicle concentration is constant throughout the lens. When we trace rays within this structure, a modified Monte Carlo's scattering algorithm determines first which scatterer (cluster or vesicle) may scatters the ray and then, if the ray is scattered, the direction of the scattered rays. Tracing a great number (107-108) rays through the system we may have an accurate depiction of the forward scattering of light by our lens model.

## 3. Numerical simulation.

In order to test our model we have performed a numerical simulation in ZOS based on the actual experimental set-up used by Van der Berg and Ijspeert [8] for measuring "in vitro" the angular distribution of the scattered light by a whole human lens. The simulated set-up can be seen in Fig 4a). Following Van der Berg and Ijspeert, we have located the human lens within a cylindrical holder containing a solution with refractive index close to that of the water. The holder is illuminated by a point source with a divergence of around 3 deg radius. We have simulated the mobile detector used by Van der Berg *et al*, with a ZOS polar detector. In the non-sequential mode of ZOS, a detector is a surface, which may be divided in pixels, used to compute radiometric magnitudes such as flux, intensity, etc, by adding the energy of the rays that passes through this surface. A polar detector -see in Fig. 4a)- is then a semispherical surface which can compute the angular distribution of the light scattered by the lens, more specifically the radiant intensity. The curvature radius of the polar detector is 550 mm the same as the distance between the lens holder and mobile photodetector used in Van der Berg and Ijspeert experiment [8]. Fig. 4b) shows an enlarged image of the layered structure of our lens model. All the simulations have been computed by tracing $10^8$ rays in non-sequential mode using an AMD® Ryzen Threadripper 3.4 GHz computer with 32 Gb of RAM memory using the version 16.5 of ZOS.

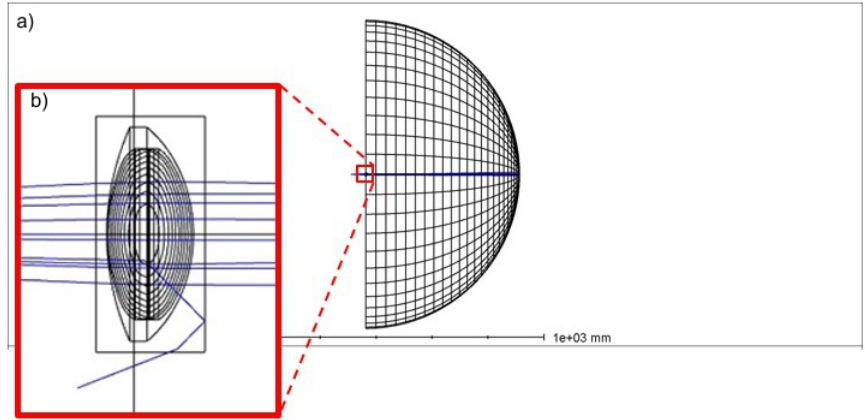

Fig. 4 a) Simulation of Van der Berg and Ijspeert experiment in Zemax using our layered human lens model. b) Amplified image of a section of the human lens placed into the cylindrical holder. The different layers of the human lens can be readily seen in this image. A set of rays are also traced showing a scattering event in the inner layers of the lens. Compare with the drawing of the original experiment (see Fig 1 of reference [8])

We have performed first an analysis of the influence of the algorithm parameters (cluster radius, inter-cluster distance and vesicle concentration on the angular distribution of the scattered light given by the function $\log_{10} I(\theta)$, being $I(\theta)$ the radiant intensity measured by the polar detector and averaged for the azimuthal angle. In order to compute the transmittance of the lens we have done the experiment disabling the scattering calculation, so ZOS traces the rays without using our bulk scattering DLL. In these conditions, an estimation of the transmittance is the quotient [8]

$$T = \frac{\int_0^{\pi/2} I(\theta) \sin\theta \, d\theta}{\int_0^{\pi/2} F(\theta) \sin\theta \, d\theta} \tag{11}$$

where $F(\theta)$ is the radiant intensity obtained when no scattering is present.

In Fig. 5 we have plotted the angular distribution of the scattered light computed for three different concentrations of vesicles ($N = 500$ mm-3, $N = 5000$ mm-3 and $N = 50000$ mm-3)for three different values of $r_0$: 20, 60 and 100 nm, respectively. For all these plots the cluster distance was $d_0 = 1200$ nm and we selected the 80 years old model. Vesicle concentration are the ones given in Costello's paper for a normal (500 mm-3) and cataractous (5000 mm-3) lens and the third concentration (50000 mm-3) would represent a highly cataractous lens.

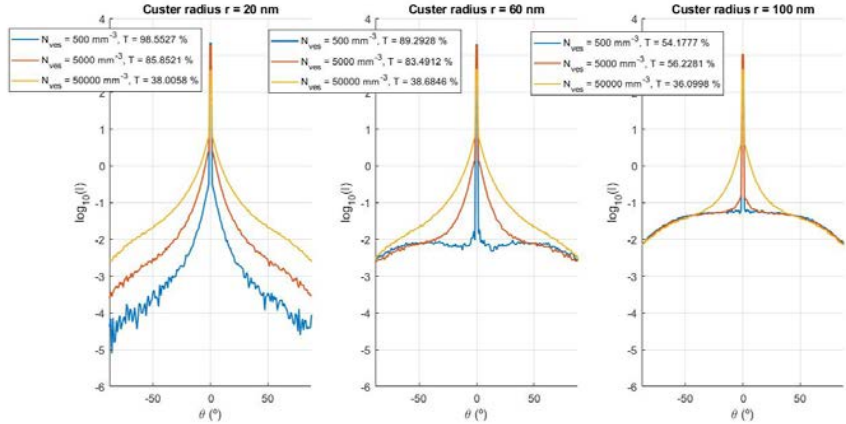

Fig 5 Angular distribution of the scattered light for different values of the vesicle concentration. The average cluster radius $r_0$ are: a) $r_0 = 20$, b) $r_0 = 60$ and c) $r_0 = 100$ nm. For all these plots, the cluster distance $d_0$ is 1200 nm. The age of the lens is 80 years.

In Fig 5a) we can see that, for a small cluster radius, all the scattering is due to the vesicles so the dispersion is greater for lower angles (which is characteristic of the scattering produced by particles whose asymmetry coefficient g is close to 1). Notice that the transmittance values obtained for the low and medium concentration of vesicles, 98.6 % and 85.6%, respectively, corresponds with the relative scattering reported by Costello et al [11] which values of 2.1% for a normal and 14.6% for a cataractous lens. In Figs 5b) and 5c) we can see that when the cluster radius becomes greater, light is scattered more evenly in all directions, particularly for small vesicle concentration. We can see that transmittance diminishes with cluster radius and vesicle concentration.

Finally, we have performed a non-linear least squares fitting of our model to the data obtained by Van der Berg and Ijspeert a young (24 year old) subject with clear lens and an elderly (80 yerar old) subject with cataracts [8]. To fit these data we have used the layered models of these lenses depicted in Fig. 3 following the gradient index model of Navarro *et al* [41]. The fitting parameters are: average cluster radius, $r_0$, cluster distance at $z = 0$, $d_0$, vesicle concentration, $N_v$, and a normalization parameter, $I_N$, which has been introduced to take into account the fact that Van der Berg data were normalized [8]. We have employed a global optimization algorithm programed in Matlab who can control ZOS files through an API (application programming interface). Fig. 6 shows the results obtained for both a normal and a cataractous eye.

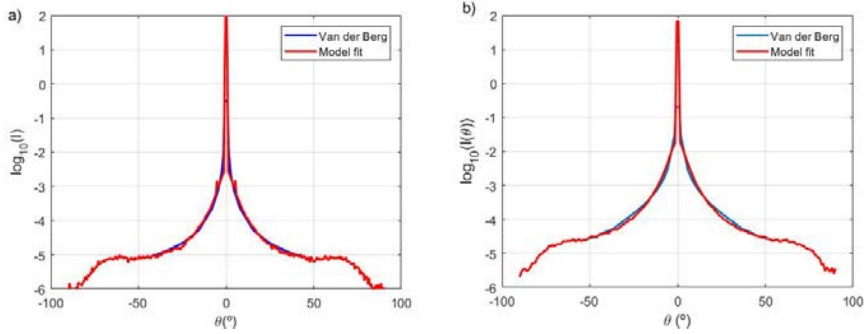

Fig 6. Angular distribution of the normalized radiant intensity corresponding to the data measured by Van der Berg and Ijspeert [8] (blue) and the fitting of our model to this data (red) for both a) a normal 24 year old eye, and b) a cataractous 80 year old eye. The plots show a good fitting between experimental data and our model.

The values of the fitting parameters are $r_0 = 31.4\ nm, d_0 = 1560\ nm, N_v = 300$ mm$^{-3}$, and $I_N = 65$ for the normal eye, and $r_0 = 29\ nm, d_0 = 1040\ nm, N_v = 1025$ mm$^{-3}$, and $I_N = 31.6$ for the cataractous one. The value of cluster diameter for both normal and cataractous lens are closed to the 50 nm reported by Benedek [19] from a purely theoretical calculation, and it is compatible with the size of the amorphous and fibril aggregates shown in [25]. However, the cluster radius obtained is larger than the size of cluster reported by [24] from "in-vitro" UV induced protein aggregation. The difference between the clusters of normal and cataractous eyes is not their size (it is slightly larger for normal eye), but their concentration, as clusters are more closely packed for the cataractous eye than for the normal one. Besides, the vesicle concentration for cataractous eye is three times that of the normal eye with values of the same order of magnitude of that of Costello *et al*. Thus, for a cataractous eye, we have found a concentration of 1025 vesicles/mm$^3$ against 4071 vesicles/mm$^3$ reported by Costello *et al* [11] and, for a normal eye, we get 300 vesicles/mm$^3$ against 556 vesicles/mm$^3$. It is clear that in our case, both clusters and vesicles contributed to the scattering and this may explain the reduction in vesicle concentration.

### 4. Conclusions

In this work we present an inhomogeneous model for characterizing numerically the scattering properties of the human eye. Our model is based in three points: 1) a rigorous FEM simulation of the properties of particles (protein clusters and vesicles) with the actual dimensions and refractive index as reported by the literature, 2) a two particle Monte Carlo bulk scattering ray tracing, with the proper modifications for taking into account the dependence of the scattering parameters with the local protein concentration, and 3) a layered model of the human lens which takes into account the inhomogeneous distribution of the particles within the human lens. Notice that our model considers two sources of inhomogeneity in the human lens: the different protein concentration (and, thus, refractive index) within the human lens, and the variation in size and number of particles along different zones of the lens.

We have tested our model by simulating numerically a classical experiment [8] in which the angular distribution of the radiant intensity scattered by the human lens was measured. By varying the model parameters: radius of protein cluster, distance between clusters and vesicle concentration along the different layers of the human lens, we have been able to calculate the relative effect of these parameters in the amount of light scattered and the shape of the scattering distribution.

Finally we have fitted our model to the experimental results reported in reference [8] showing a good agreement between them for both a normal and a cataractous eye. The values of the fitting parameters obtained are feasible and they show how the cataractous eye present a small increase in the cluster protein concentration while tripling the concentration of vesicles.

Given the characteristics of our model, it might be used as a workbench for analyzing the effect of different configuration of the particles responsible for the ocular scattering in a realistic way. It also may be helpful in establishing new diagnostic techniques based on the scattering of light by the human lens particles.

**Funding**

This work has been financed by the European Fund for Regional Development (EFRD-FEDER, EU) and the Spanish Government's Agencia Estatal de Investigación (AEI) belonging to Ministerio de Economia y Competitividad (MINECO) through the grant DPI2016-75272-R.

**Acknowledgements**

We wish to thank Dr. Javier Alda for his useful comments and his help in the elaboration of this manuscript.

**Disclosures**

The authors declare that there are no conflicts of interest related to this article